\documentclass[prl,twocolumn,showpacs,preprintnumbers,amsmath,amssymb]{revtex4}

\usepackage{graphicx}% Include figure files
\usepackage{dcolumn}% Align table columns on decimal point
\usepackage{bm}% bold math
\usepackage[bookmarks = true, pdfpagemode = None, pdfstartview = FitH,colorlinks = true, citecolor = black, linkcolor = black, urlcolor = black]{hyperref}
\usepackage{url}

\newcommand{\ket}[1]{\left\vert{#1}\right\rangle}
\begin{document}

\preprint{DWAVE/COUPLER-FINAL}
\title{Sign-- and Magnitude--Tunable Coupler for Superconducting Flux Qubits}
\author{R.~Harris}
\email{rharris@dwavesys.com}
\author{A.J.~Berkley}
\author{M.W.~Johnson}
\author{P.~Bunyk}
\author{S.~Govorkov}
\author{M.C.~Thom}
\author{S.~Uchaikin}
\author{A.B.~Wilson}
\author{J.~Chung}
\author{E.~Holtham}
\author{J.D.~Biamonte}
\author{A.Yu.~Smirnov}
\author{M.H.S.~Amin}
\author{Alec~Maassen van den Brink}
\altaffiliation{Currently at Dept. of Physics, Simon Fraser University, Burnaby, Canada}
\email{alec@riken.jp}
\affiliation{%
D-Wave Systems Inc., 100-4401 Still Creek Dr., Burnaby, BC V5C 6G9, Canada
}%
\homepage{www.dwavesys.com}

\date{\today}

\begin{abstract}

We experimentally confirm the functionality of a coupling element for flux-based superconducting qubits, with a coupling strength $J$ whose sign and magnitude can be tuned {\it in situ}. To measure the effective \nobreak$J$, the groundstate of a coupled two-qubit system has been mapped as a function of the local magnetic fields applied to each qubit. The state of the system is determined by directly reading out the individual qubits while tunneling is suppressed. These measurements demonstrate that $J$ can be tuned from antiferromagnetic through zero to ferromagnetic.

\end{abstract}

\maketitle

%\section{Introduction}

One approach to building useful quantum processors is to abandon the gate model in favor of physics-inspired models, such as adiabatic quantum computation (AQC) and its variants~\cite{Farhi}. In AQC one adiabatically evolves the processor from the groundstate of an initial Hamiltonian \nobreak${\cal H}_{\rm i}$ (chosen such that this groundstate can be readily prepared) to the groundstate of a final~${\cal H}_{\rm f}$, encoding the solution to the problem of interest.  For example, an Ising magnet is a physical system that could be harnessed for this purpose~\cite{Aeppli}.  AQC is known to be computationally equivalent to the circuit model of quantum computation~\cite{aharonov2004}. Realistic AQC architectures exist~\cite{Kaminsky} that could be used to produce accurate approximate solutions to NP-complete problems~\cite{GareyJohnson}. One such architecture, based on superconducting electronics~\cite{screviews}, requires devices that couple qubit pairs with \emph{in situ} programmable coupling magnitude and sign~\cite{patents}. In this letter we demonstrate such a coupler between two superconducting flux qubits.

\begin{figure}[ht]
\includegraphics[bb=0 0 570 350, width=3.4in]{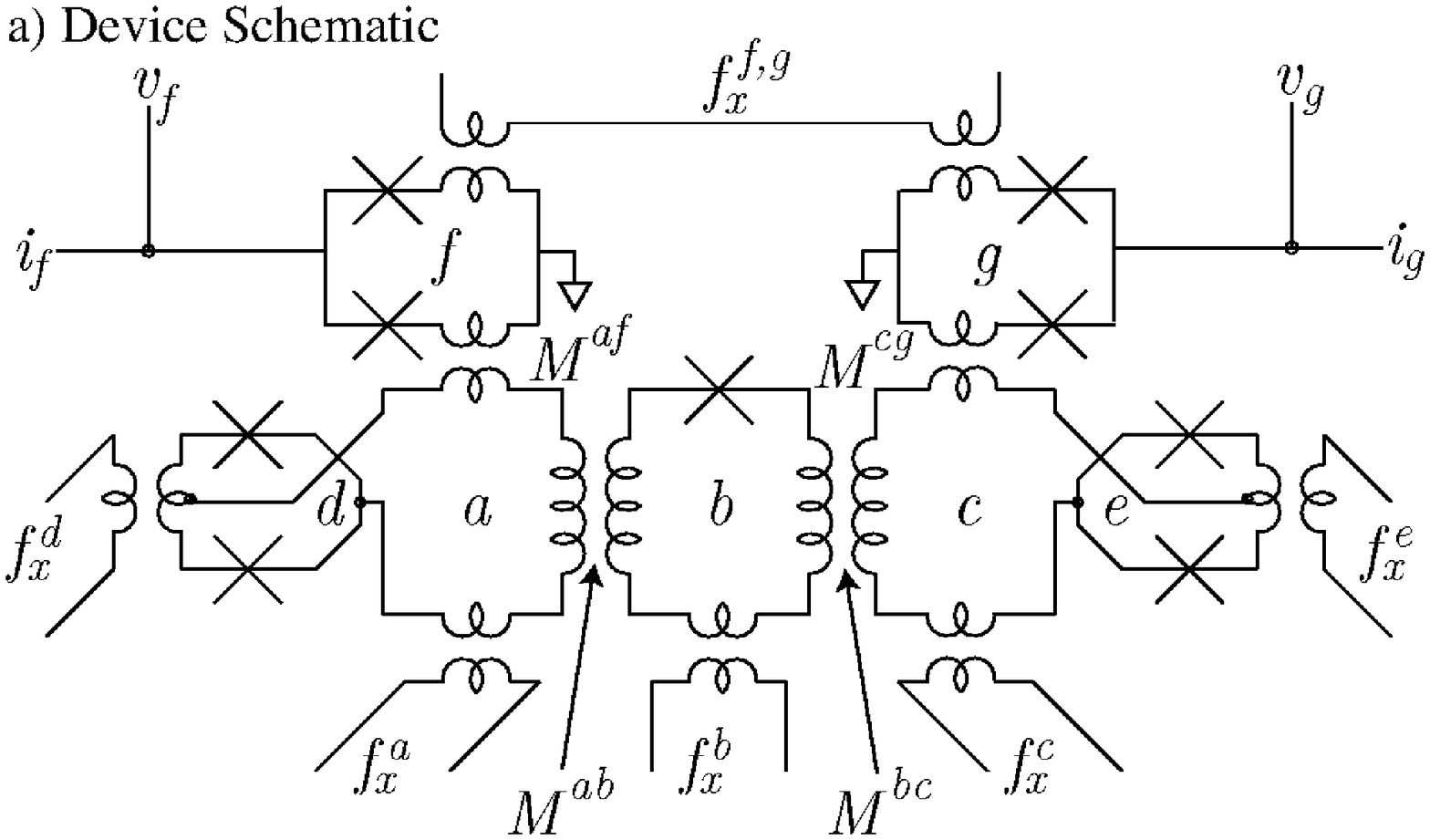}\\
\includegraphics[width=3.38in]{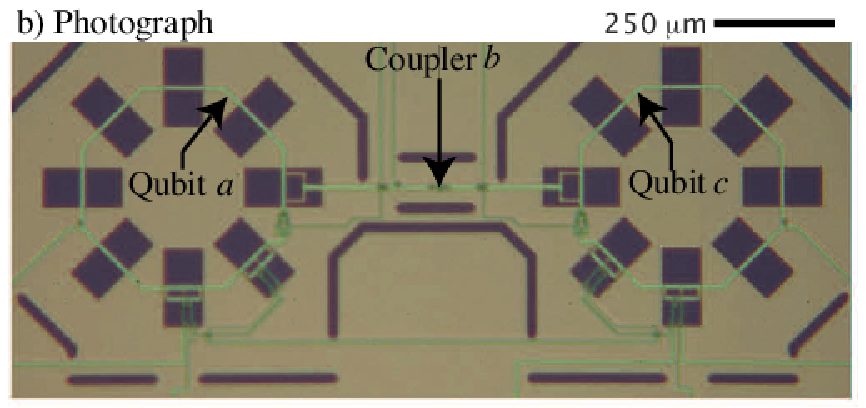}\\
\includegraphics[width=3.38in]{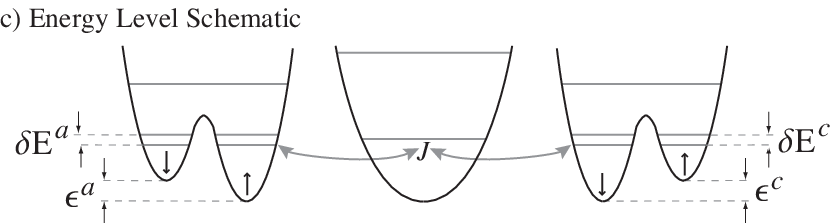}
\caption{a)~Device schematic: SQUID loops $i$, mutual inductances $M^{ij}$, flux bias lines $f_x^i$, current bias lines $i_i$ and voltage taps $v_i$ as indicated (see text). b)~Photograph of device. c)~Energetics of the coupled qubit system.  Qubits are represented as bistable potentials with pseudospins as indicated.  Local biases 
$\epsilon^i$ are applied to each qubit.  The lowest lying qubit states are  
separated by $\delta E^i=\sqrt{(\epsilon^i)^2+(\Delta^i)^2}$.
The coupler is represented as a monostable potential that mediates the interaction $J$ [see Eq.~(\ref{eqn:fourlevel})].}\label{fig:deviceschematic}
\end{figure}

The qubit design used is a bistable rf-SQUID, magnetically biased near its degeneracy point by an external flux $\Phi_x\approx\frac{1}{2}\Phi_0$ ($\Phi_0=h/2e$ is the flux quantum) \cite{Leggett1985,fluxqb}. The qubits are denoted by $a$, $c$ in Fig.~\ref{fig:deviceschematic}, with bias controls $f_x^a$,~$f_x^c$ ($f_x^i\equiv\Phi_x^i/\Phi_0$). Compound Josephson junction (CJJ) loops $d$, $e$ with biases $f_x^d$, $f_x^e$ are employed to tune the critical currents $I_c^i$~\cite{cjjref}. For readout, the qubits are inductively coupled to their own dedicated dc-SQUIDs $f$, $g$~\nobreak\cite{Cosmelli}, biased to points of high flux sensitivity via a shared control~$f_x^{f\!,g}$. The switching currents were measured by ramping the bias currents $i_f$, $i_g$ and monitoring the voltages at points $v_f$,~$v_g$.  Tunable interqubit coupling is mediated via a monostable rf-SQUID $b$ with separate flux bias $f_x^b$~\nobreak\cite{AMvdBcoupler}.  This approach and that of Refs.~\cite{condmat0605588} and \cite{ClarkeDCSQUIDCoupler} are recent demonstrations of \emph{in situ} sign-- and magnitude--tunable coupling elements between flux qubits.

%Loops $a$ and $c$ are \mbox{rf-SQUID} qubits with flux biases $f_x^a$ and~$f_x^c$. %Loop $b$ is the rf-SQUID coupler with bias~$f_x^b$. Biases $f_x^d$ and $f_x^e$ act %on qubit compound junction loops $d$ and $e$. Loops $f$ and $g$ are \mbox{dc-
%SQUIDs} with current bias lines $i_f$ and $i_g$ and shared flux bias line $f_x^{f\! ,g}$. %Voltages are monitored at points $v_f$ and $v_g$.

%\section{Coupler principle}

We couple the qubits to the coupler loop via transformers leading to mutual inductances $M^{ab}$, $M^{bc}$ so that the qubits are influenced the coupler's persistent current
$I_p^b = I_c ^b\sin(2\pi f^b)$, not the control flux~$\Phi_x^b$. Here $f^b$, the total flux in loop $b$, is a function of $f_x^b$~\cite{ClassicalRFSQUID}. The persistent current $I_p^a$ in qubit $a$ alters the flux applied to $b$ by $I_p^aM^{ab}\ll\Phi_0$.  This additional flux alters $I_p^b$ by an amount $\delta I_p^b\approx I_p^aM^{ab}\chi^b$, where the susceptibility $\chi^b\equiv dI_p^b/d\Phi_x^b$. $\delta I_p^b$ in turn alters the flux applied to qubit~$c$ by $\delta I_p^bM^{bc}$, causing an interaction of energy
\begin{equation}
  \label{eqn:J1}
  J = M^{ab}M^{bc} \chi^b I_{p}^aI_{p}^c\;,
\end{equation}

\noindent where $I_p^c$ is the persistent current in qubit $c$.  Thus the coupler mediates an effective qubit--qubit interaction. Since $I_p^b$ and therefore $\chi^b$ are periodic in $f_x^b$ with alternating sign, this interaction can be ferromagentic (FM,$\chi^b<\nobreak0$), antiferromagnetic (AFM,$\chi^b>0$), or zero.

The energy spacing between the ground and first excited states of the coupler was designed to be much higher ($\gg10\,$GHz) than typical qubit splittings.  As such the coupler is expected to remain in its groundstate and the effective low-energy qubit--coupler--qubit Hamiltonian~\cite{AMvdBcoupler} can be written as follows:
\begin{equation}\label{eqn:fourlevel}
  {\cal H} = -\sum_{q=a,c}\bigl(\epsilon^q \sigma_{z}^{(q)} + \Delta^q \sigma_{x}^{(q)}\bigr)
   + J(f_{x}^b) \sigma_{z}^{(a)} \sigma_{z}^{(c)}\;,
\end{equation}
where $\sigma_{x}^{(q)}$ and $\sigma_{z}^{(q)}$ are Pauli matrices for qubit $q$.  Here $\epsilon^q$ and $\Delta^q$ represent the energy bias and tunnel splitting for the individual qubits, as indicated in Fig.~\ref{fig:deviceschematic}c. Note that $\epsilon^q\propto f_x^qI_p^q$ and $J$ [Eq.~(\ref{eqn:J1})] scale with $I_p^a$ and $I_p^c$. In turn, $I_p^q$ and $\Delta^q$ are functions of $f_x^d$ and $f_x^e$ for $q=a$ and $c$, respectively.  Tuning these latter two flux biases provides a means of annealing the system from a quantum regime, where there is appreciable tunneling, to the classical regime where $\Delta^q\rightarrow 0$.

%\section{Device Fabrication}

The circuit was fabricated on an oxidized Si wafer using a Nb trilayer process with
wiring layers isolated by a sputtered SiO$_2$ dielectric~\cite{jpl}. The qubit inductance was designed to be $500\,$pH. The total parallel capacitance of each CJJ (single junction area $A=0.7\times0.7\,\mu$m$^2$) was designed to be $C^q=20\,$fF. Fitting of macroscopic resonant tunneling peaks \cite{tla} for a similar device at large $I_p^q$ yielded $C^q=(33\pm3)\,$fF.  The discrepancy may be due to capacitive loading of the junctions by nearby wiring. The unsuppressed $I_c^q$ of each CJJ was designed to be $2.5\,\mu$A. The single-junction ($A=0.6\times0.6\,\mu$m$^2$) coupler has designed inductance $L^b=240\,$pH and $I_c^b=~0.9\,\mu$A, giving $\beta^b\equiv2\pi L^b I_c^b/\Phi_0=0.65$. Fits to the measured $M^{ab}I_p^b$ and $M^{bc}I_p^b$  (explained below) yielded $\beta^b=~0.63\pm~0.02$. The mutual inductances were designed to be {\nobreak $M^{ab}=~M^{bc}\equiv~M^{qb}=~25\,$pH.} Measurements on similar breakout structures yielded $M^{qb}=(28\pm 2)\,$pH.

\begin{figure}
\includegraphics[width=3.4in]{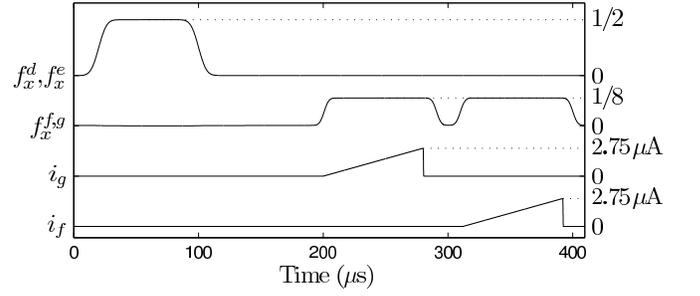}
\caption{Bias sequence. CJJ biases $f_x^d$, $f_x^e$ are raised to $\frac{1}{2}$, held for $60\,\mu$s, and then returned to zero over $20\,\mu$s. Readout dc-SQUID flux bias $f_x^{f\! ,g}$ is held at zero during qubit evolution and raised to its operating point during the current bias ramps on $i_f$, $i_g$. Control lines $f_x^a$, $f_x^b$, and $f_x^c$ are held constant throughout this procedure.}\label{fig:ControlSequence}
\end{figure}

Qubit states were inferred from the bias currents $i_f$, $i_g$ required to switch the dc-SQUIDs into the voltage state. These elements had designed maximum critical currents of 2.5$\,\mu$A for $f_x^{f\! ,g}=0$ ($A=0.7\times0.7\,\mu$m$^2$). The bias currents were ramped linearly from zero to 2.75$\,\mu$A in $80\,\mu$s (see Fig.~\ref{fig:ControlSequence}) and voltages monitored at points \linebreak $v_f$ and $v_g$. Voltage trigger thresholds were set at $\sim$1.5$\,$mV and monitored with $10\,$ns timing resolution. The two persistent current states of each qubit generated a flux difference of $(16\pm1)\,$m$\Phi_0$ in their respective detector with the width of the switching distributions limited by the detector sensitivity of $1.2\,$m$\Phi_0$. Using the design value $M^{af}=M^{cg}=11\,$pH, we estimate $2I_p^q=(3.0\pm0.2)\,\mu$A when $f_x^d=f_x^e=0$.

The device was mounted in an Al box in a dilution refrigerator with base temperature below $10\,$mK. Two coaxial cryo\-perm shields surrounded the sample area. Battery powered current control electronics were located in an rf shielded room together with the fridge, and interfaced though fiber optics with a PC outside. All control lines had discrete element filters at the $1\,$K stage and mixing chamber as well as copper powder filters attached to the latter. All flux bias couplings were weak, minimizing the dissipation introduced into the qubits from the environment via the wiring in the fridge.

In order to account for cross coupling between the flux bias controls we measured the mutual inductance matrix between lines \emph{in situ} by tracking features with known period $\Phi_0$ in each of the qubits and the coupler.  Better design can avoid linear crosstalk but not nonlinear qubit biasing due to $I_p^b$~\cite{AMvdBcoupler}. To calibrate the latter we suppressed $I_p^q$ of one of the qubits $q$ and measured the state of qubit $q'\neq q$ versus $f_x^b$ and~$f_x^{q'}$. The degenerate bias $\Phi_x^{q'}$ now follows from $M^{q'b}I_p^b+\Phi_x^{q'}=\frac{1}{2}\Phi_0$. Fitting these data determined $\beta^b$ as noted above.   In addition such measurements were used to generate $M^{q'b}\chi^b$ which can then be used to predict the functional form of $J$ (Eq.~\ref{eqn:J1}) up to a prefactor $M^{qb}I_p^aI_p^c$.

%\section{Measurement Technique}

We measured $J$ by mapping the interacting groundstate versus $\epsilon^a$, $\epsilon^c$ in the limit $\Delta^q\ll |J|$ where Eq.~(\ref{eqn:fourlevel}) is nearly a classical Ising Hamiltonian with eigenstates $\ket{ac}=\ket{\uparrow\uparrow}$, $\ket{\uparrow\downarrow}$, $\ket{\downarrow\uparrow}$, and~$\ket{\downarrow\downarrow}$. The groundstate is found via an annealing procedure: $\Delta^q$ are first increased by biasing the CJJs at $f_x^{d,e}=\frac{1}{2}$ to render the qubits monostable, thus initializing them in a known groundstate.  Thereafter $\Delta^q$ are slowly lowered back to their minima at $f_x^{d,e}=0$ (see Fig.~\ref{fig:ControlSequence}).  The system will remain in the groundstate during the lowering of $\Delta^q$ until the rate of evolution of ${\cal{H}}$ exceeds a limit set by the physics of a Landau--Zener (LZ) transition~\cite{LandauZener}. The measured fluxes represent the groundstate of Eq.~(\ref{eqn:fourlevel}) at the last instance before this transition. Thus according to Eq.~(\ref{eqn:J1}), the observed $J$ will depend on particular values of $I_p^a(f_x^d)$ and $I_p^c(f_x^e)$, to be determined experimentally, which can be less than the maximum values observed at $f_x^d=f_x^e=0$.

%\section{Results}

\begin{figure}[t]
\begin{tabular}{cc}
\includegraphics[width=1.65in]{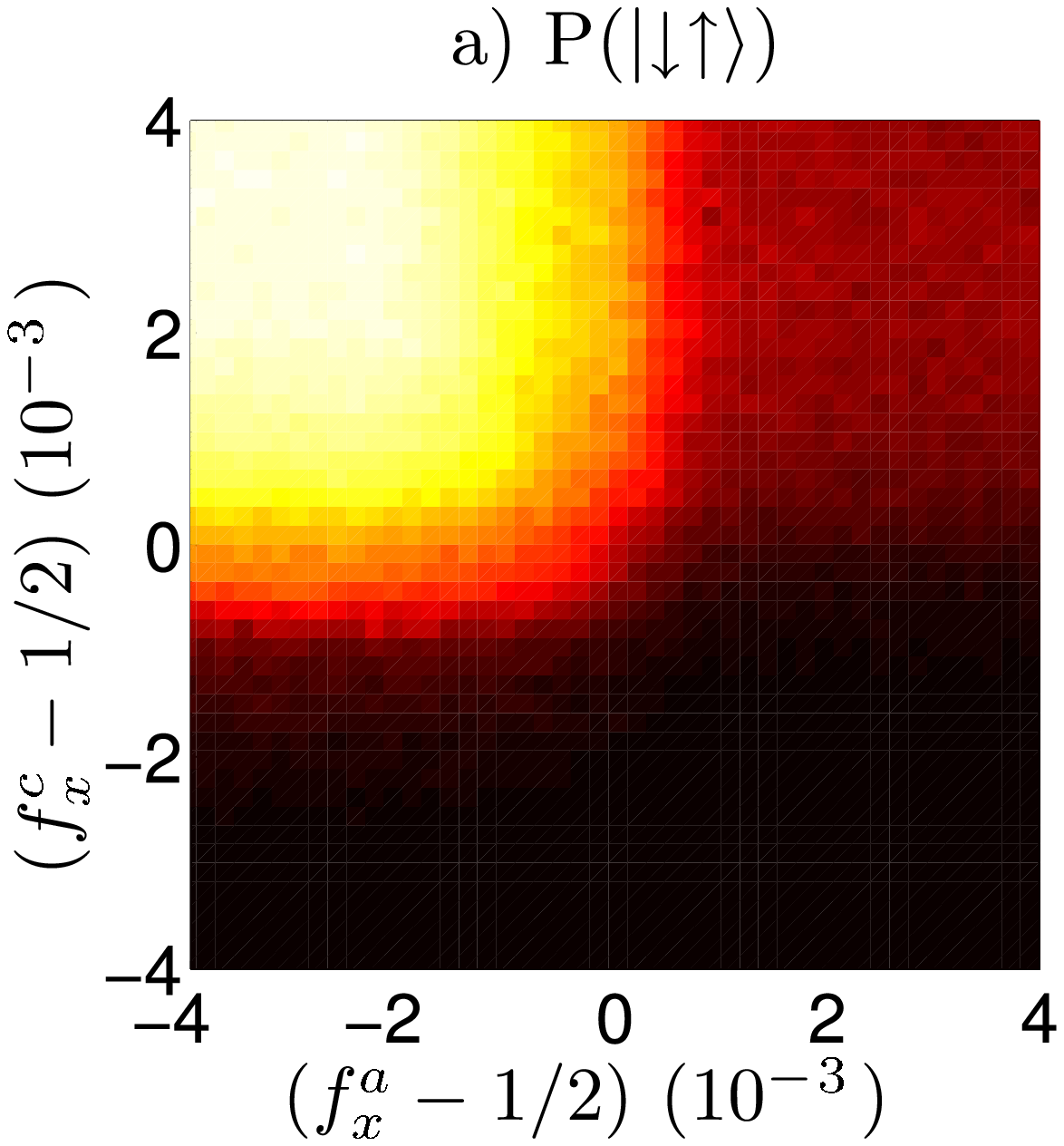} &
\includegraphics[width=1.65in]{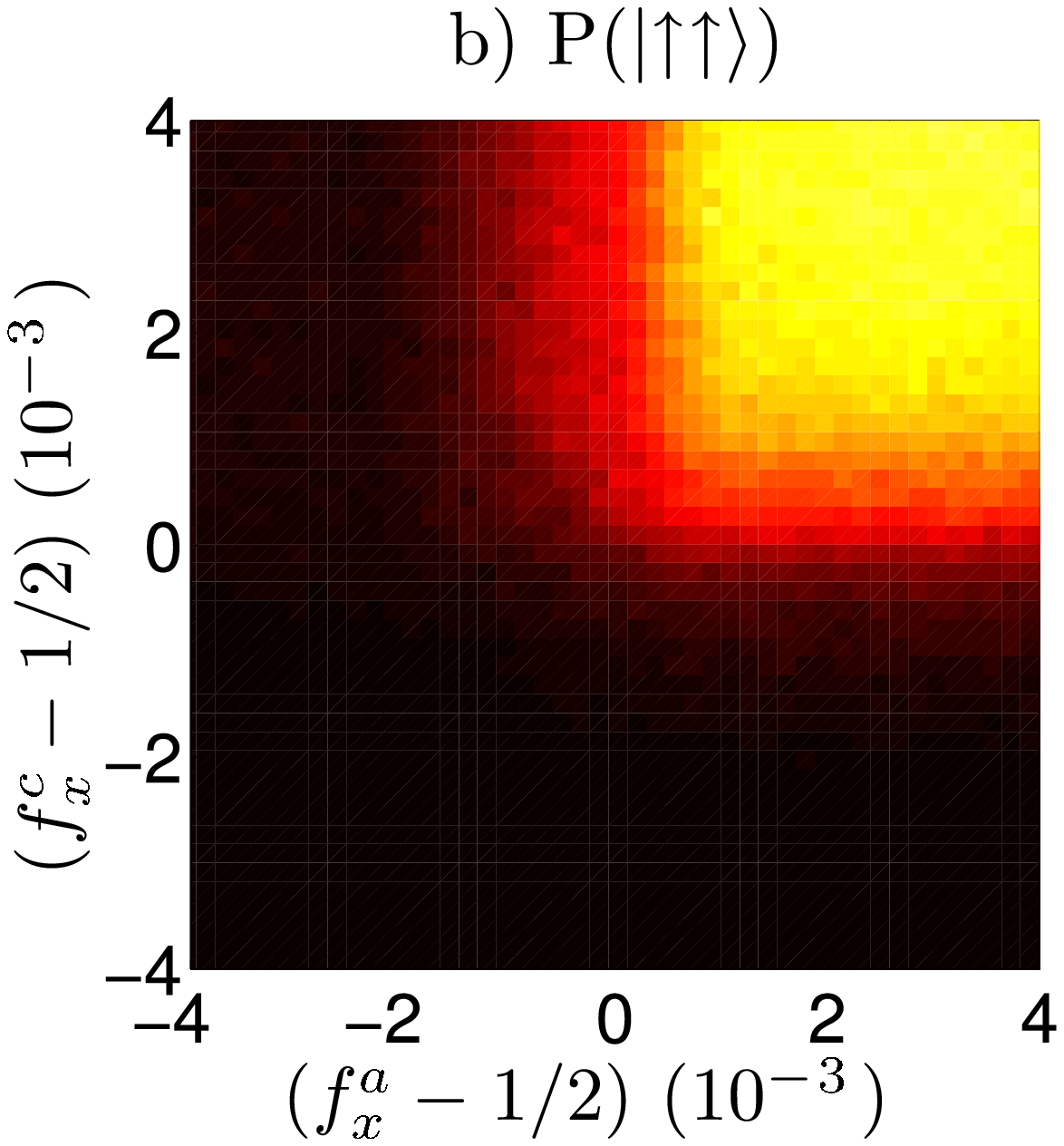} \\
\includegraphics[width=1.65in]{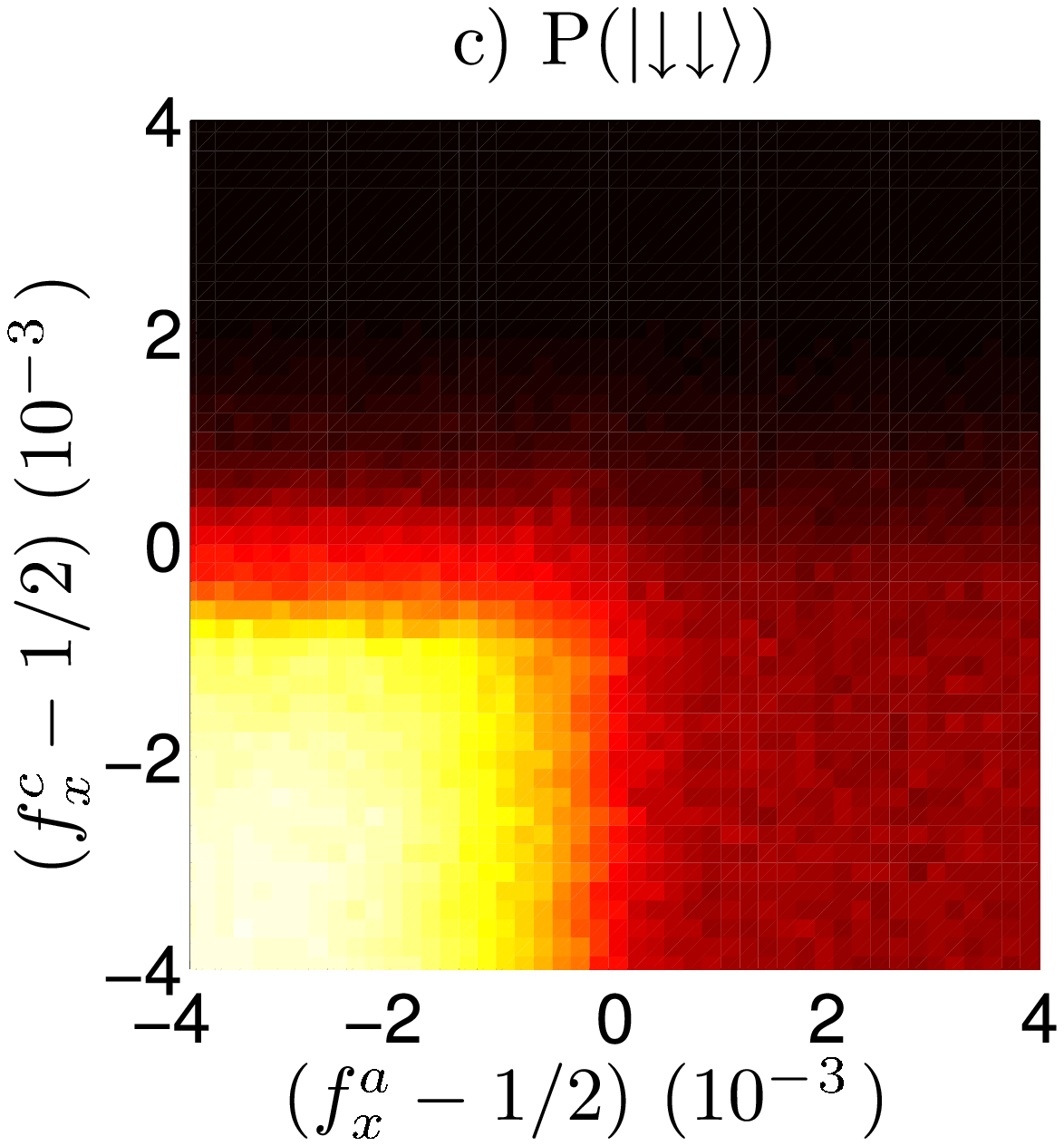} &
\includegraphics[width=1.65in]{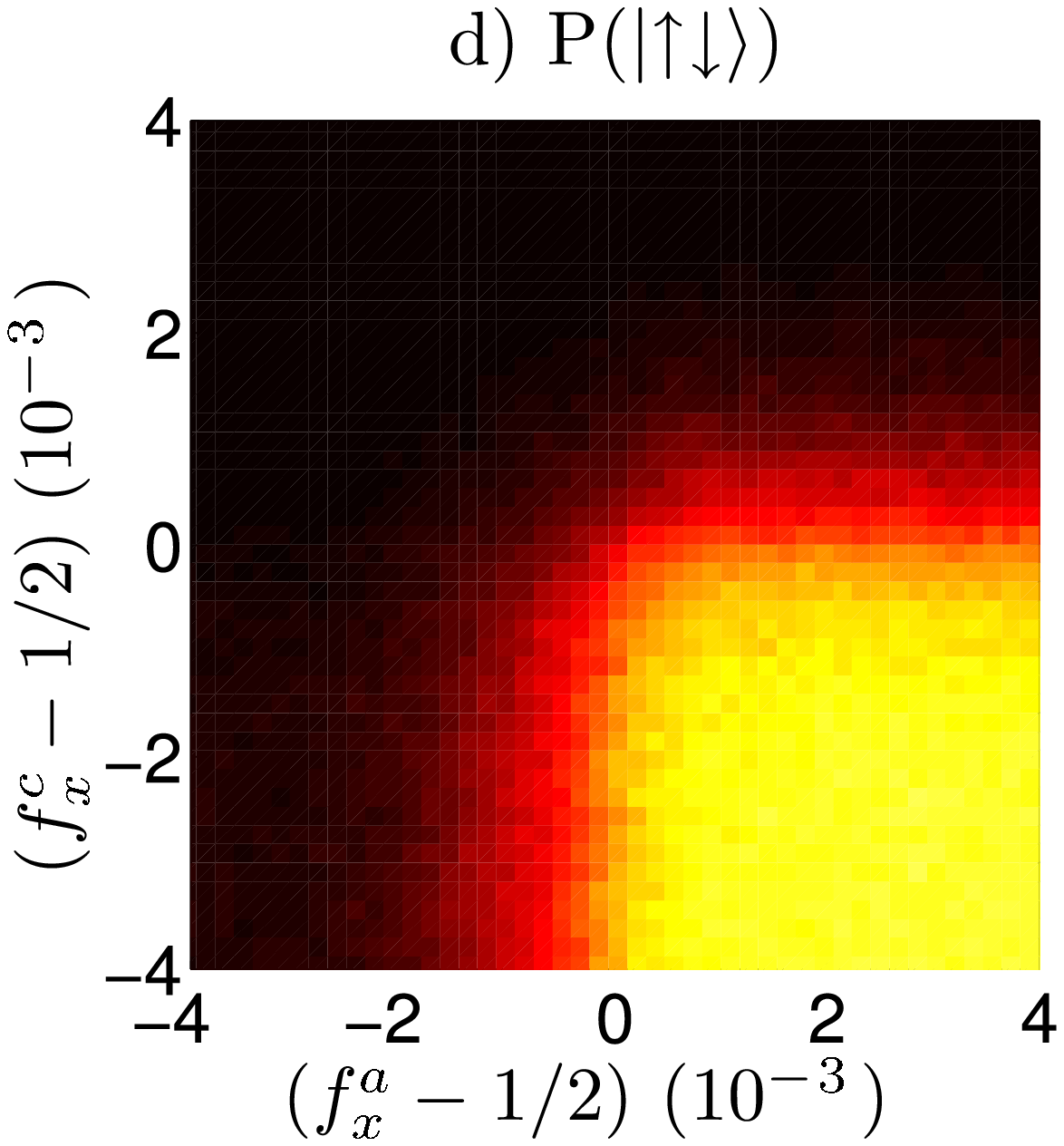} \\
\end{tabular}
\\
\includegraphics[width=3.3in]{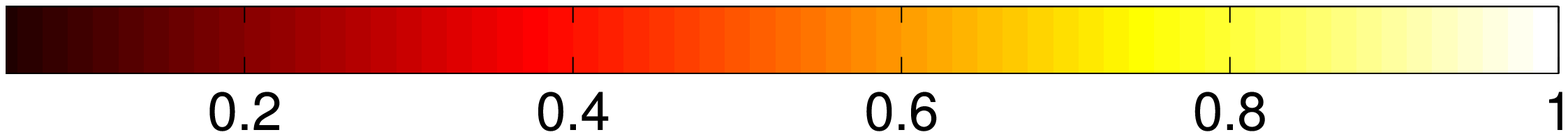}
\caption{Measured probabilities for each of the four possible flux states for small $\Delta_q$, at coupler flux bias $f_x^b=-0.03$.}
\label{fig:SampleStateMapPlot}
\end{figure}

Coupler performance was demonstrated at three values of~$f_x^b$, corresponding to maximum AFM ($f_{x}^b= 0$), zero [$f_x^b=f_{J=0}^b\in (0,\frac{1}{2})$] and maximum FM ($f_x^b=\frac{1}{2}$) coupling. At each bias point, the above annealing sequence was repeated 1024 times and the probabilities $P(\ket{\uparrow\uparrow})$ etc.\ of finding the four classical flux states were determined. For this, direct readout of individual qubits in our design is a key advantage. As an example, Fig.~\ref{fig:SampleStateMapPlot} displays $P(\ket{ac})$ for $f_x^b=-0.03$. The flux states observed near the corners $(f_x^a - \frac{1}{2},f_x^c - \frac{1}{2})=(\pm4,\pm4)\times 10^{-3}$ agree with those expected from Eq.~(\ref{eqn:fourlevel}) for $\epsilon^a,\epsilon^c\gg J$. While $P(\ket{\downarrow\uparrow})$ (Fig.~\ref{fig:SampleStateMapPlot}a) and $P(\ket{\downarrow\downarrow})$ (Fig.~\ref{fig:SampleStateMapPlot}c) exceed 95\% at the top and bottom left corners respectively, it appears that $P(\ket{\uparrow\uparrow})$ (Fig.~\ref{fig:SampleStateMapPlot}b) and $P(\ket{\uparrow\downarrow})$ (Fig.~\ref{fig:SampleStateMapPlot}d) only reach 
$\sim$85\%. Further investigation revealed that this was due to measurement crosstalk from the readout of qubit~$c$.   Note that qubit~$c$ is read first in the bias sequence depicted in Fig.~\ref{fig:ControlSequence}.  It was verified that qubit~$a$ could be read without interference if the $i_g$ control pulse were absent.  However, switching of the readout SQUID $g$ into the voltage state was found to disturb the state of qubit $c$, which in turn influenced the state of qubit $a$ prior to its measurement.  Observations indicate that the $\ket{\uparrow\downarrow}$ and $\ket{\uparrow\uparrow}$ states were most susceptible to corruption on this particular device.

\begin{figure*}[ht]
\begin{tabular}{ccc}
\includegraphics[width=2.325in]{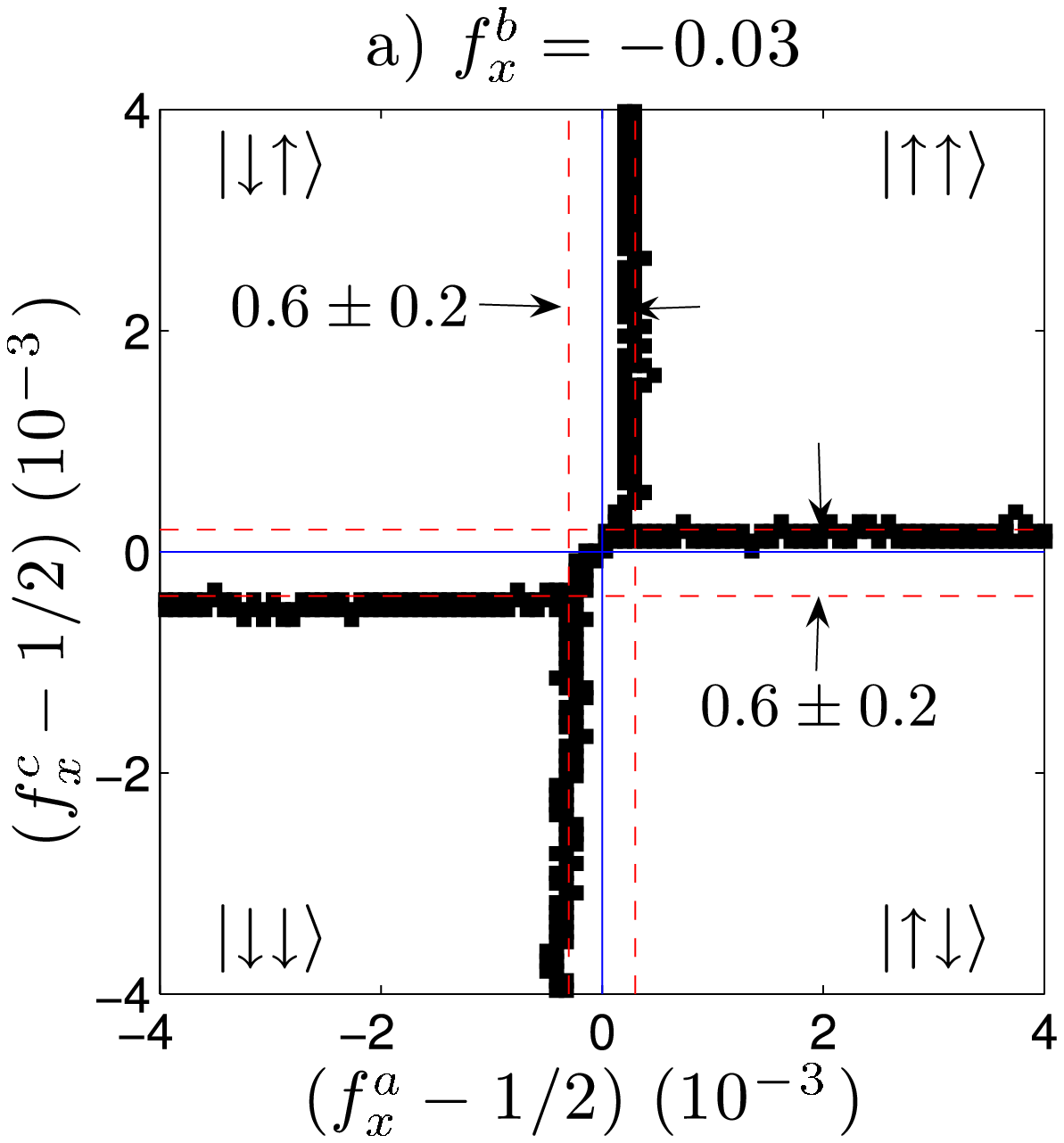} &
\includegraphics[width=2.325in]{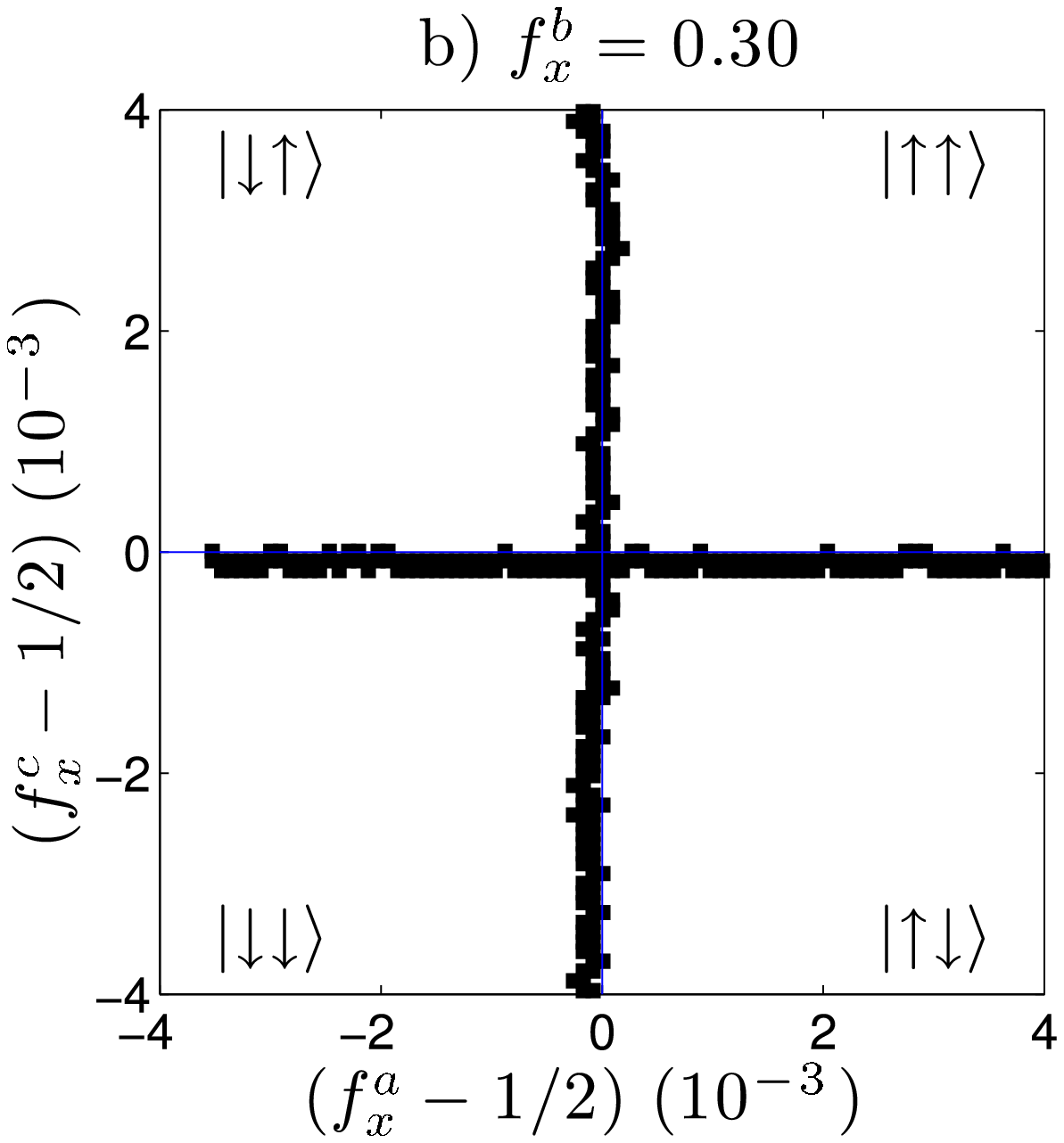} &
\includegraphics[width=2.325in]{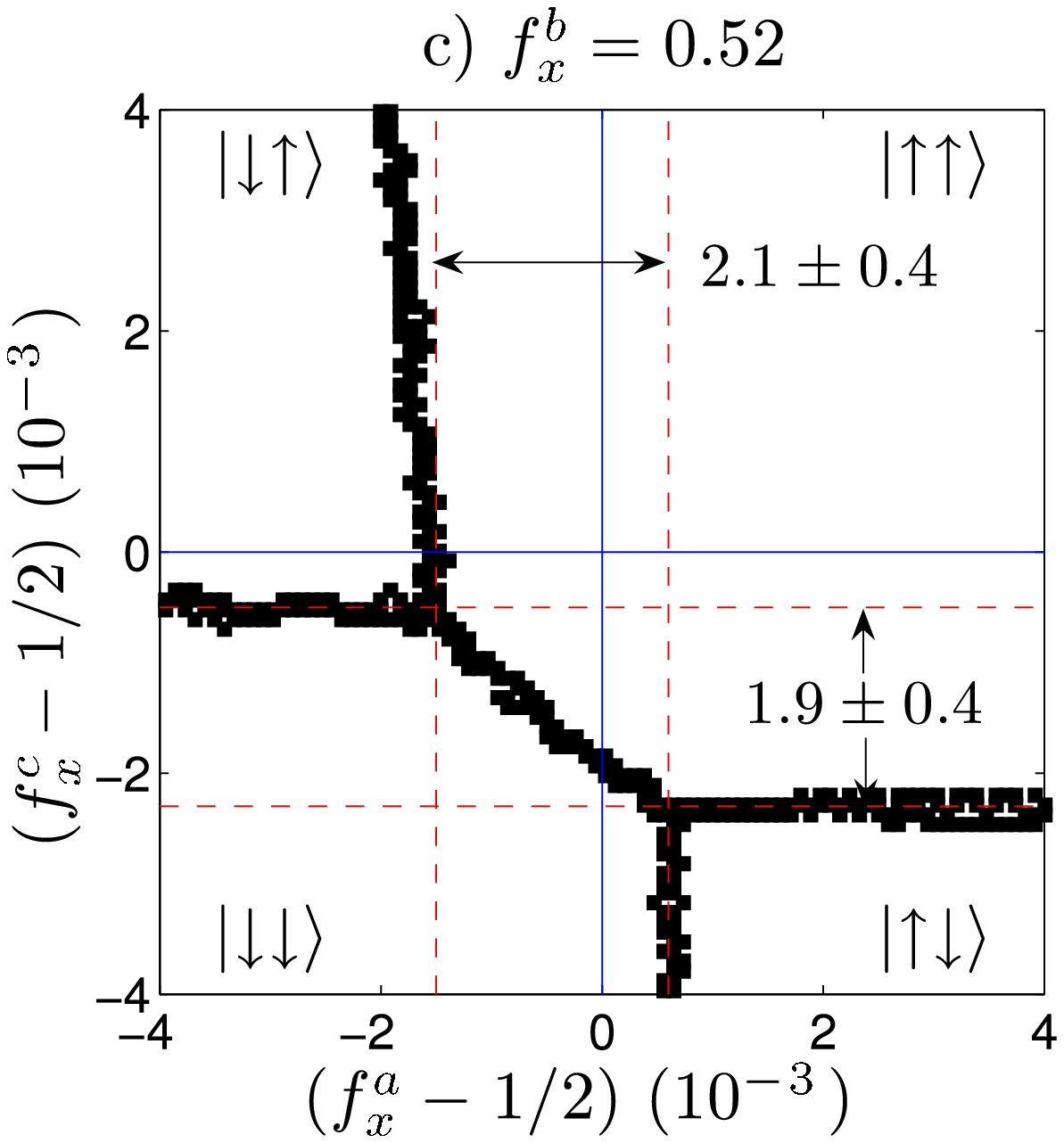}
\end{tabular}
\caption{Groundstate stability diagrams: a) AFM coupling, b) zero coupling and c) FM coupling.  The calibrated degeneracy lines for the individual qubits and extreme values of coupled flux are indicated by solid and dashed lines, respectively.}
\label{fig:BoundaryPlot}
\end{figure*}

Groundstate stability diagrams were generated by determining which flux state occurs with the highest probability at each point $(f_x^a,f_x^c)$. The experimentally determined boundaries are shown in black in Fig.~\ref{fig:BoundaryPlot}. In Fig.~\ref{fig:BoundaryPlot}a, $f_x^b= -0.03$, and the AFM states $\ket{\uparrow\downarrow}$ and $\ket{\downarrow\uparrow}$ dominate the groundstate map. One clearly observes a boundary between these two states, which only occurs if $J>0$. This demonstrates that our device can provide an \textit{AFM qubit--qubit interaction}. Next, for $f_x^b=0.30$, Fig.~\ref{fig:BoundaryPlot}b shows the critical case of \textit{zero coupling} ($J=0$): the state of each qubit is independent of the flux applied to the other, resulting in a cross structure with the fourfold degeneracy point at $(f_x^a,f_x^c)=(\frac{1}{2},\frac{1}{2})$. Finally, for $f_x^b=0.52$ (Fig.~\ref{fig:BoundaryPlot}c), the FM regions $\ket{\downarrow\downarrow}$ and~$\ket{\uparrow\uparrow}$ are enhanced. The new boundary between them is present only if $J<0$, indicating a \textit{FM qubit--qubit interaction}. Ideally in all cases the boundaries should have intersected the $J=0$ degeneracy point. However, the transitions appear to have translated by $(f_x^a,f_x^c)\sim(0,-0.1)\,$m$\Phi_0$ in the AFM case and by $(f_x^a,f_x^c)\sim(-0.5,-1.5)\,$m$\Phi_0$ in the FM case.  Further investigation revealed that this discrepancy was due to the readout crosstalk problem noted earlier and was aggravated by strong interqubit coupling in the FM case, which then altered the population statistics.

We determine $J$ from either the displacement between the $\ket{\downarrow\uparrow}$:$\ket{\uparrow\uparrow}$ and $\ket{\downarrow\downarrow}$:$\ket{\uparrow\downarrow}$ boundaries (vertical dashed lines in Fig.~\ref{fig:BoundaryPlot}) or between the $\ket{\uparrow\uparrow}$:$\ket{\uparrow\downarrow}$ and $\ket{\downarrow\uparrow}$:$\ket{\downarrow\downarrow}$ boundaries (horizontal dashed lines). According to Eq.~(\ref{eqn:fourlevel}), in the limit $\Delta^q/|J|\ll1$, the former should yield a net flux equal to $2|J|/I_p^a$ and the latter $2|J|/I_p^c$. The observed horizontal and vertical displacements are equal to within experimental error for both the FM [mean $(0.6\pm0.2)\times 10^{-3}$] and AFM [mean $(2.0\pm0.4)\times 10^{-3}$] cases. One concludes that the qubits and transformers are reasonably symmetric, i.e., $I_p^a\approx I_p^c\equiv I_p^q$. 

\begin{figure}[t]
\includegraphics[width=3.5in]{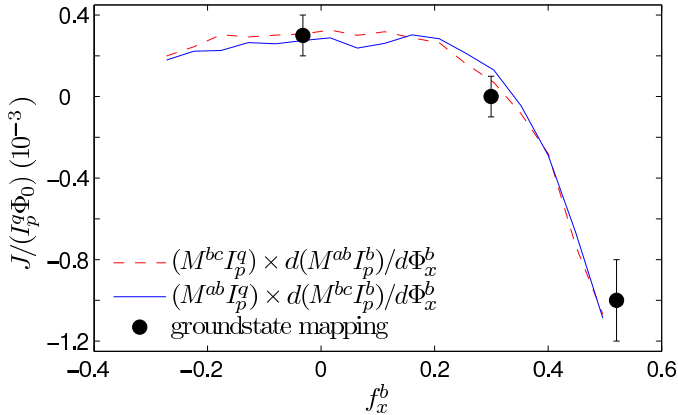}
\caption{Coupling strength (in units of coupled flux) versus $f_x^b$ as obtained from groundstate mapping and from $dM^{qb}I_p^q/d\Phi_x^b$ scaled by $M^{qb}I_p^q$ with $I_p^q=(0.5\pm0.1)\,\mu$A [see Eq.~(\ref{eqn:J1})].}
\label{fig:CouplingStrengthPlot}
\end{figure}

The measured coupled flux values $J(f_x^b)/(I_p^q \Phi_0)$ are shown in Fig.~\ref{fig:CouplingStrengthPlot} together with the measured $d(M^{qb}I_{p}^b)/d\Phi_x^b$ scaled by $M^{qb}I_p^b$, where $I_p^q$ is used as a free parameter.  The observed amount of coupled flux is in reasonable agreement with the predicted form (Eq.~[\ref{eqn:J1}]) if the qubit currents are $\sim\nobreak I_p^q=(0.5\pm0.1)\,\mu$A at the last moment before the LZ transition.  Decreasing the magnitude of the ramp rate of $f_x^d$ and $f_x^e$ should allow the system to evolve more slowly as $\Delta^q$ decreases, thus shifting the inevitable LZ transition to a larger $I_p^q$.  As such, this measure of $I_p^q$ at which adiabatic evolution appears to terminate is not due to a fundamental limitation.  

The key result of this study is a clear demonstration of a \emph{sign and magnitude} tunable coupler between two superconducting flux qubits. While originally designed for a specific quantum processor architecture, this device may prove useful for various superconducting electronics applications, both quantum and classical, requiring \emph{in situ} tunability. Further, this work demonstrates the utility of a multi-qubit readout technique, directly measuring individual qubits, which is scalable to larger qubit numbers and may be of importance in future practical quantum processors.

%\begin{acknowledgments}
We thank S.~Han, E.~Ladizinsky, J.~Hilton, G.~Rose, A.~Tcaciuc,  F.~Cioata, 
A.O. Niskanen, and Y.~Nakamura for useful discussions. Samples were fabricated by B.~Bumble, A.~Fung, and A.~Kleinsasser of the Microelectronics Laboratory of the Jet Propulsion Laboratory,  operated by the California Institute of Technology under a contract with the National Aeronautics and Space Administration.
%\end{acknowledgments}

\end{document}